\documentclass[graybox]{svmult}

\usepackage{mathptmx,helvet,courier}
\usepackage{type1cm}
\usepackage{amsmath, amsfonts, amssymb}
\usepackage{graphicx,color}
\usepackage{newtxtext}
\usepackage{newtxmath}
\usepackage[bottom]{footmisc}

\usepackage{makeidx}
\usepackage{multicol}
\makeindex

\usepackage{lineno}
\modulolinenumbers[5]
\begin{document}

\title*{GPU-based parallel simulations of the Gatenby-Gawlinski model with anisotropic, heterogeneous acid diffusion}
\titlerunning{Simulation on GPUs of anisotropic, heterogeneous acid-mediated tumor invasion}
%%% http://inspire.norceresearch.no/fvca9/-overview
%%% https://ocs.springer.com/prom/home/FVCAIX

\author{Corrado Mascia, Donato Pera and Chiara Simeoni}

\institute{Corrado Mascia \at Dipartimento di Matematica G. Castelnuovo, Sapienza Universit\`a di Roma,\\
piazzale Aldo Moro 2 - 00185 Roma - Italy\\
\email{corrado.mascia@uniroma1.it}
\and Donato Pera \at Dipartimento di Ingegneria e Scienze dell'Informazione e Matematica,\\
Universit\`a degli studi dell'Aquila, via Vetoio (snc) - 67100 Coppito (AQ) - Italy\\
\email{donato.pera@univaq.it}
\and Chiara Simeoni (corresponding author) 
\at Laboratoire de Math\'ematiques J.A. Dieudonn\'e CNRS UMR 7351,\\
Universit\'e C\^ote D'Azur, Parc Valrose - 06108 Nice Cedex 2 - France\\
\email{chiara.simeoni@univ-cotedazur.fr}}

\maketitle

%%%%%%%%%%%%%%%%%%%%%%%%%%%%%%

\abstract{We introduce a variant of the Gatenby-Gawlinski model for acid-mediated tumor invasion, accounting for anisotropic and heterogeneous diffusion of the lactic acid across the surrounding healthy tissues. Numerical simulations are performed for two-dimensional data by employing finite volume schemes on staggered Cartesian grids, and parallel implementation through the modern CUDA GPUs technology is considered. The effectiveness of such approach is proven by reproducing biologically relevant results like the formation of propagating fronts and the emergence of an interstitial gap between normal and cancerous cells, which is driven by the pH lowering strategy and depends significantly on the diffusion rates. By means of a performance analysis of the serial and parallel execution protocols, we infer that exploiting highly parallel GPU-based computing devices allows to rehabilitate finite volume schemes on regularly-shaped meshes, together with explicit time discretization, for complex applications to interface diffusion problems of invasive processes.}
\medskip

\medskip

%%% https://mathscinet.ams.org/msc/msc2010.html

%%%%%%%%%%%%%%%%%%%%%%%%%%%%%%

\section{Biological context and mathematical modeling}
\label{sec:introduction}

The phenomenological framework concerning the \textit{Warburg effect} relies on the experimental results achieved by Otto Warburg~\cite{Warb1956} in the 1920s, which essentially prove that cancer cells metabolism leans on anaerobic glycolysis for adenosine triphosphate (ATP) production, regardless of the available oxygen amount, hence causing lactic acid fermentation. %(as shown in Figure~\ref{figurextra12})
Such a behavior leads to the \textit{acid-mediated invasion hypothesis}~\cite{SmGa2008}, whose key point consists in assuming that acidification induced by an excess of lactic acid sets up a toxic microenvironment for normal cells and, consequently, favors malignant cells spreading.
%%% more references~\cite{Warb1930,GaGa2006,SmGa2005}
%%% as a matter of fact, oxygen turns out to be the principle/main resource to allow normal cells performing more efficient glucose metabolism, due to the best yield of adenosine triphosphate (ATP) production by using oxidative phosphorylation (anaerobic versus aerobic metabolism)

%\begin{figure}%[b]
%\sidecaption%[t]
%\includegraphics[scale=.25]{figurextra1}
%%\includegraphics[scale=.35]{figurextra2}
%\caption{Glycolytic metabolism of tumor cells versus oxidative phosphorylation of normal cells for adenosine triphosphate (ATP) production.}
%\label{figurextra12}
%\end{figure}

%modeling acid-mediated tumor invasion: the tumor phenotype is characterized by a metabolism of glycolytic type resulting in an increased acidity -- invasive tumors switch to glycolytic metabolism and the prevailing phenotype (in the Darwinian-like mutation process from normal to tumor cells) is acid resistant (apoptosis threshold for normal cells pH=7.1 (Casciari et al., 1992) whilst for tumor cells pH=6.8 (Dairkee et al., 1995)) -- although the glycolytic metabolism is inefficient from the energetic point of view, it provides a decisive advantage in the invasion process by raising the acidity of the environment; moreover, aggressive phenotypes are characterized by low oxygen consumption, high proliferation rate, little or no adhesion (mesenchymal phenotype) and high haptotaxis coefficients

From the point of view of mathematical modeling, these qualitative statements are properly framed by a system of reaction-diffusion equations known as the \textit{Gatenby-Gawlinski model}~\cite{GaGa1996}, which describes tumor cells proliferation and progression inside the local healthy tissue through the destructive effect of superfluous acidity, thus focusing on a stage of biological development in which the carcinogenesis has already occurred.
%%% the situation considered in~\cite{GaGa1996,GaGa2003} could be extended to consider the occurrence of necrotic cores~\cite{SmGa2005} as already mentioned in~\cite{FoHo1973} and the glucose dynamics (see~\cite{BiFa2009}, for instance)
The crucial feature revealed also by analytical investigations is the emergence of propagating fronts connecting equilibrium states, but the available literature is currently restricted to almost only one-dimensional problems~\cite{FaHe2009,McGa2014,DaHe2018}.
%%% namely one-dimensional problems and radially symmetric configurations
%%% more references~\cite{GaGa2003,TaTe2016}
%%% the existence of traveling waves connecting equilibrium states...
Multidimensional configurations have been recently considered in~\cite{MoSi2019} to support experimental observations, by performing two- and three-dimensional finite element simulations exploiting the COMSOL Multiphysics$^{\textregistered}$ environment~\cite{CoMu2020} together with ParaView utilities~\cite{PaVi2020} for post-processing three-dimensional graphical results.

\begin{figure}%[b]
\sidecaption%[t]
\qquad \includegraphics[scale=.65]{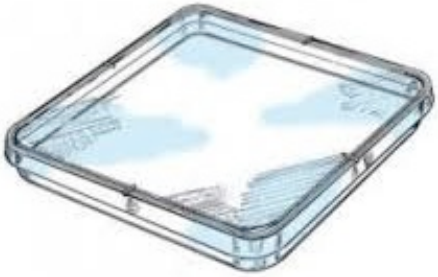}
\qquad \includegraphics[scale=.30]{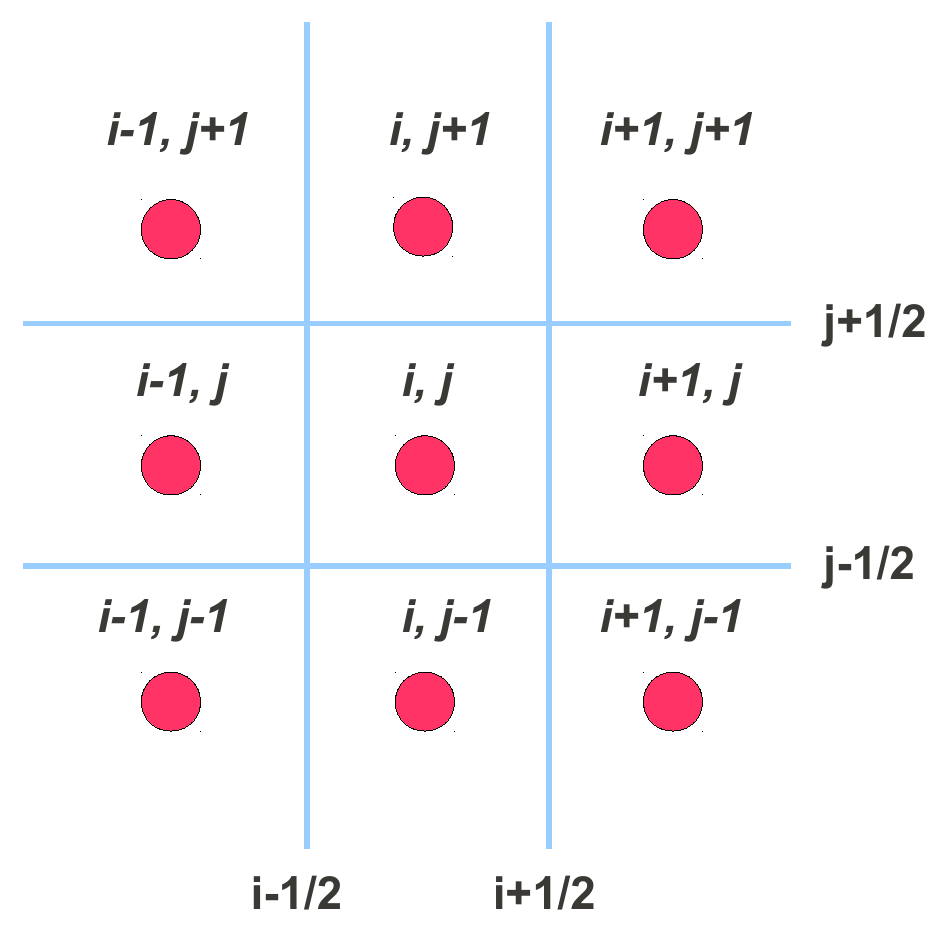}
\caption{Squared \textit{Petri dish} for in vitro experiments (left) and staggered Cartesian mesh (right)}
\label{figure1}
\end{figure}

In this article, we introduce a biologically relevant variant of the Gatenby-Gawlinski system, which is displayed in nondimensionalized form as follows:
\begin{equation}
\label{eqn:multisystem}
\left\{ \begin{aligned}
\partial_t U &= U (1-U) - d\,U W\\
\partial_t \!V &= r V (1-V) + \alpha \nabla\!\cdot \!\big[ (1-U) \nabla V \big]\\
\partial_t \!W &= c (V-W) + \nabla\!\cdot \!\big( \mathbb{A} \nabla W \big)
\end{aligned} \right.
\end{equation}
%%% which is defined for some time interval $(0,T)$ and a suitable domain $\Omega \subset \mathbb{R}^m$, $m=2$ or $3$, so that the differential operators are $\nabla=(\partial / \partial x_1, ..., \partial / \partial x_m)$ and $\nabla^{2}=\sum_{i=1}^m\partial^{2} / \partial x_i^2$, with $m=2$ or $3$
for $0\leq t \leq T$ and a suitable domain $\Omega \subset \mathbb{R}^2$, so that the first order differential operator is denoted by $\nabla=(\partial_x,\,\partial_y)$ and $\mathbb{A}$ is an anisotropic, heterogeneous (positive definite) diffusion tensor, namely
\begin{equation}
\label{eqn:anisodiff}
\mathbb{A}(x,y) = \begin{bmatrix}
a(x,y) & b(x,y) \\ b(x,y) & c(x,y)
\end{bmatrix}\,.
\end{equation}
%%% the structure of the diffusion matrix is often determined by phenomenological observations about the problemÕs background: in particular, anisotropic and heterogeneous operators occur when the mass diffusion is faster in some directions than others (like transportation over oriented networks), and variable rates are applied proportionally to local features of the media~\cite{BaJo2002,SaSa2002}
%%% moreover, the principles of dynamical processes~\cite{Crank1975,Tham2011} dictate the choice of positive definite symmetric tensors, thus we focus on the models characterized by...
For the sake of convenience, the problem is defined on the square of sides $L$ corresponding to a two-dimensional support typically employed in biology for cell cultures (see Figure~\ref{figure1}(left)), and Neumann boundary conditions are taken into account.

\noindent The dynamical variables of system~\eqref{eqn:multisystem} are the scaled unknowns $U(t;x,y)$, $V(t;x,y)$ and $W(t;x,y)$ which represent the normal cells density, the tumor cells density and the extracellular lactic acid concentration in excess, respectively. The cellular populations obey a logistic growth with normalized carrying capacities, and $r$ is the relative growth rate of malignant species. The degradation of healthy tissue by means of the lactic acid is based on the \textit{law of mass action}~\cite{Murr2002} in the form of a second-order kinetic reaction with (adimensionalized) velocity $d$, such a term in the first equation translating the outcome of the pH lowering strategy adopted by cancer cells to enhance their invasive processes.
%%% the (mediated) chemical action of the tumor is an invasive process driven by pH lowering: the destructive effect of H+ ions on healthy tissue translates into the removal of normal cells, so that the (adimesionalized) constant $d$ is a death rate proportional to lactic acid concentration
%%% the healthy and cancerous cells are viewed as two species following logistic growth with mutual competition, and the interaction between the tumor and the vascular system is neglected
Another remarkable feature is the density-limited, and possibly degenerate, diffusion of the tumor cells in the second equation, as a consequence of the defense mechanism of confinement by the host tissue experimentally noticed in various contexts~\cite{SwRa2009}, with $\alpha$ being the unconstrained diffusion constant.
%%% reflecting the fact that a healthy tissue may spatially constrain a tumor unless shrunk~\cite{GaGa1996,SaMa1995} and $\alpha$ is the unconstrained diffusion constant when the local healthy tissue has already been degraded
Finally, in the third equation, the parameter $c$ encloses both the rate of production for the tumor-associated acidity and a physiological reabsorption rate.
%%% H+ ions are supposed to be produced at constant rate by tumor cells (acid production proportional to $V$) and, for example, the lactic acid could be cleared by vasculature

The novelty of system~\eqref{eqn:multisystem} with respect to the original Gatenby-Gawlinski model is that the (excess of) lactic acid diffuses across the microenvironment according to an anisotropic and heterogeneous tensor~\eqref{eqn:anisodiff}, which is suggested by the intrinsic geometrical and/or structural complexity of the extracellular matrix~\cite{Pain2009,DyGr2016}, among other factors. On the other hand, the chemical aggression at the expenses of the local healthy tissue usually includes protease reactions inducing lysis of the extracellular matrix~\cite{ChAn2003}, and therefore the assumption of an isotropic diffusion for cancer cells is also justified.
%%% more references~\cite{GeCh2008}
%%% the extracellular matrix (ECM) is the tissue surrounding the tumor, and typically degraded by the cancer cells; it is a complex mixture of macromolecules, some of which are believed to play structural roles while others are involved in cells adhesion, spreading and motility
%%% typically, the early growth stages of a primary solid tumor are merely a consequence of the successive divisions of their initial constituents; however, when a critical size is reached, further growth requires aggressive action from the tumor on its surrounding tissue: this is done in several ways, one of which consists in co-opting blood vessels to provide oxygen and nutrients to the expanding colony (to this end, tumors are known to produce angiogenic substances that stimulate and attract blood vessels); on the other hand, they may also release products which favor the degradation of neighboring tissues, thus facilitating the progression of tumor cells through such tissues (the subsequent destruction of the host tissues under this invasive process is a major ingredient in the morbidity associated with tumor diseases, and the invasive process is actually determined by the ability of the tumor to degrade the surrounding ECM by production of the MDEs)
The environment-driven rate of invasion of malignant species is a widely reported phenomenon~\cite{SwAl2000,ChHi2007,PaHi2013} and we aim at initiating a systematic study of the effect of anisotropic, heterogeneous diffusion operators also within the context of acid-mediated tumor invasion, for which reliable and efficient numerical simulations are necessary because of the unavailability of theoretical results.

%%%%%%%%%%%%%%%%%%%%%%%%%%%%%%

\section{Finite volume discretization on staggered grids}
\label{sec:numerics}

The numerical simulation of system~\eqref{eqn:multisystem} relies on finite volume schemes for the spatial operators and the explicit Euler method for time integration, to take advantage from desirable properties like conservativity and high-order accuracy~\cite{LeVe2002}.
%%% finite volume schemes on staggered grids are particularly well-suited for the transport and/or diffusion fluxes of interface problems, with $x_{i+1/2}$ and $y_{j+1/2}$ located at the initial interfaces

In practical applications, since the diffusion tensor~\eqref{eqn:anisodiff} and problem data are derived from the experimental setting, the solution is normally sampled at the nodes of a fixed two-dimensional Cartesian mesh (see Figure~\ref{figure1}(right)). We consider the domain $\Omega=[0,1]\times[0,1]$ which is partitioned into cells of characteristic sizes $\Delta x$ and $\Delta y$, so that the grid points are given by $(x_i,y_j)=(i\,\Delta x,\,j \Delta y)$, $i,j=1,2,...$
%%% the physical domain $\Omega$ is discretized as follows: given an integer $N$, we denote by $\Delta x=\Delta y=\frac1{N}$ the uniform spatial step-size and, accordingly, we define the grid $\Omega_N=\left\{ (x_i,y_j)\in \Omega\,; x_i=i\,\Delta x\,, y_j=j\,\Delta y\,, i,j=0,1,...,N \right\}$
\begin{remark}
As already mentioned in~\cite{BeAf1996}, even for computational domains with a complex geometry, using regularly-shaped Cartesian embedded grids generally makes the numerical methods efficient~\cite{BeHe2003,LuFa2016} and indeed easier to implement in parallel compared with boundary fitted structured or unstructured meshes.
%%% more references~\cite{TaSh2010}
\end{remark}

\noindent For the time discretization, starting from $t_0=0$ we proceed by selecting a nonuniform step-size $\Delta t_k$ which is computed after each iteration in order to ensure numerical stability of the finite volume scheme, so that $t_{k+1}=t_k+\Delta t_k$, $k=0,1,...$\\
%%% in order to fulfill the (modified) CFL-condition
Finally, we prescribe auxiliary interfacial coordinates, namely $x_{i+\frac12}\!=\!\dfrac{x_i+x_{i+1}}2$ and $y_{j+\frac12}\!=\!\dfrac{y_j+y_{j+1}}2$, to construct dual cells $C_{ij}\!=\!\big(x_{i-\frac12},x_{i+\frac12}\big)\!\times\!\big(y_{j-\frac12},y_{j+\frac12}\big)$ where the numerical unknowns are defined at each time frame as cell-centered approximations
\begin{equation*}
%\label{eqn:cellaverage}
U_{ij}^k \approx \frac1{\big| C_{ij} \big|} \int_{C_{ij}}\hspace{-5pt}U(t_k;x,y)\,dx\,dy\,, \quad \big| C_{ij} \big| = \Delta x\Delta y\,,
%%% \quad V_{ij}^k \approx \frac1{\big| C_{ij} \big|} \int_{C_{ij}}\hspace{-5pt}V(t_k;x,y)\,dx\,dy\,, \quad W_{ij}^k \approx \frac1{\big| C_{ij} \big|} \int_{C_{ij}}\hspace{-5pt}W(t_k;x,y)\,dx\,dy\,,
\end{equation*}
for instance, and analogously for the other variables of system~\eqref{eqn:multisystem},
%%% for any fixed time $t$, we denote by $U_{ij}, V_{ij}, W_{ij}$ (by omitting the dependence on the $t$ variable, for simplicity) an approximation in the discrete space (i.e. an approximate value) of $\int_{C_{ij}} f(x,y)\,dx\,dy$ for $C_{ij}=\left[ x_{i-\frac12}, x_{i+\frac12} \right) \times \left[ y_{j-\frac12}, y_{j+\frac12} \right)$, then for $i,j=1,2,...,N-1$ we obtain
together with the piecewise constant projections of the diffusion rates~\eqref{eqn:anisodiff} on the staggered mesh, which are denoted by $a_{ij}$, $b_{ij}$ and $c_{ij}$, for $i,j=1,2,...$
%%% the entries of $\mathbb{A}$ are the rate functions oriented along the different space directions

The simplest form of fully discretized problem is presented as follows:
\begin{equation}
\label{eqn:multischeme}
\left\{ \begin{aligned}
U_{ij}^{k+1} &= \,U_{ij}^k + \Delta t_k U_{ij}^k \big(1-U_{ij}^k \big) - d\,\Delta t_k U_{ij}^k W_{ij}^k\\
V_{ij}^{k+1} &= \,V_{ij}^k + r\,\Delta t_k V_{ij}^k \big(1-V_{ij}^k \big) + \alpha\,\Delta t_k \,\mathbb{D}_1(V;U)^k_{ij}\\
%+ \alpha \frac{\Delta t_k}{\Delta x \Delta y} (1-U) \nabla V \big|_{\partial C_{ij}}
W_{ij}^{k+1} &= \,W_{ij}^k + c\,\Delta t_k \big(V_{ij}^k-W_{ij}^k \big) + \Delta t_k \,\mathbb{D}_2(W;\mathbb{A})^k_{ij}
%+ \frac{\Delta t_k}{\Delta x \Delta y} \mathbb{A} \nabla W \big|_{\partial C_{ij}}
\end{aligned} \right.
\end{equation}
and then this scheme is applied with Neumann-type boundary conditions.
%%% that provide the values of the unknown functions $U^k_{0j}=U^k_{1j}$, $j=1,2,...$, for example, and particular attention must be payed to the (four) corner points...
To make the notation more compact, we represent the numerical diffusion operators $\mathbb{D}_1(V;U)^k_{ij}$ and $\mathbb{D}_2(W;\mathbb{A})^k_{ij}$ in~\eqref{eqn:multischeme} by means of the two-dimensional stencils reported in Table~\ref{tab:tumorstencil} and Table~\ref{tab:acidstencil} for the second and third equation, respectively, where the central entries are the coefficients of $V^k_{ij}$ and $W^k_{ij}$, the right-hand side applies to $V^k_{i+1,j}$ and $W^k_{i+1,j}$, the upper side corresponds to $V^k_{i,j+1}$ and $W^k_{i,j+1}$, and similarly for all the others.
%%% the (finite volume) numerical diffusion operators (fluxes) on staggered grids are defined at the interfaces of the dual cells $C_{ij}$, where the interfacial values $a_{i+1/2,j}$ and $a_{i-1/2,j}$ are further identified through the point-values and, for example, calculated by arithmetic averages $a_{i\pm1/2,j}=\frac{a_{ij}+a_{i\pm1,j}}2$

\renewcommand{\arraystretch}{2.75}
\begin{table}
\begin{center}
\begin{tabular}{c|c|c}
$0$ & $\dfrac1{\Delta y^2} - \dfrac{U^k_{ij}+U^k_{i,j+1}}{2 \Delta y^2}$ & $0$ \\[3.75pt] \hline 
%$0$ & $\dfrac{2-U^k_{ij}-U^k_{i,j+1}}{2 \Delta y^2}$ & $0$ \\[3.75pt] \hline 
$\dfrac1{\Delta x^2} - \dfrac{U^k_{i-1,j}+U^k_{ij}}{2 \Delta x^2}\;$ 
& $\; -\dfrac2{\Delta x^2} + \dfrac{U_{i-1,j}+2\,U_{ij}+U_{i+1,j}}{2 \Delta x^2} \qquad$ 
& $\;\dfrac1{\Delta x^2} - \dfrac{U^k_{ij}+U^k_{i+1,j}}{2 \Delta x^2}$ 
\tabularnewline
& $\qquad -\dfrac2{\Delta y^2} + \dfrac{U_{i,j-1}+2\,U_{ij}+U_{i,j+1}}{2 \Delta y^2} \;$ & \\[3.75pt] \hline 
%$\dfrac{2-U^k_{i-1,j}-U^k_{ij}}{2 \Delta x^2}\;$ 
%& $\; -\dfrac{4-U_{i-1,j}-U_{ij}-U_{i+1,j}}{2 \Delta x^2}-\dfrac{4-U_{i,j-1}-U_{ij}-U_{i,j+1}}{2 \Delta y^2} \;$ 
%& $\;\dfrac{2-U^k_{ij}-U^k_{i+1,j}}{2 \Delta x^2}$ \\[3.75pt] \hline 
$0$ & $\dfrac1{\Delta y^2} - \dfrac{U^k_{i,j-1}+U^k_{ij}}{2 \Delta y^2}$ & $0$ \\[3.75pt]
%$0$ & $\dfrac{2-U^k_{i,j-1}-U^k_{ij}}{2 \Delta y^2}$ & $0$ \\[3.75pt]
\end{tabular}
\end{center}
\caption{Five-point stencil of the standard discretization for the nonlinear tumor cells diffusion.}
\label{tab:tumorstencil}
\end{table}

\renewcommand{\arraystretch}{2.5}
\begin{table}
\begin{center}
\begin{tabular}{c|c|c}
$\; -\dfrac{b_{i-1,j}+b_{i,j+1}}{4 \Delta x \Delta y} \;$ 
& $\dfrac{c_{ij}+c_{i,j+1}}{2 \Delta y^2}$ 
& $\dfrac{b_{i+1,j}+b_{i,j+1}}{4 \Delta x \Delta y}$ \\[5pt] \hline 
$\dfrac{a_{i-1,j}+a_{ij}}{2 \Delta x^2}$ 
& $\; -\dfrac{a_{i-1,j}+2\,a_{ij}+a_{i+1,j}}{2 \Delta x^2}-\dfrac{c_{i,j-1}+2\,c_{ij}+c_{i,j+1}}{2 \Delta y^2} \;$ 
& $\dfrac{a_{ij}+a_{i+1,j}}{2 \Delta x^2}$ \\[5pt] \hline 
$\dfrac{b_{i-1,j}+b_{i,j-1}}{4 \Delta x \Delta y}$ 
& $\dfrac{c_{i,j-1}+c_{ij}}{2 \Delta y^2}$ 
& $\; -\dfrac{b_{i+1,j}+b_{i,j-1}}{4 \Delta x \Delta y} \;$ \\[5pt]
\end{tabular}
\end{center}
\caption{Nine-point stencil of the standard discretization for the (linear) anisotropic, heterogeneous lactic acid dissusion.}
\label{tab:acidstencil}
\end{table}

\noindent The consistency and stability of finite volume schemes on staggered Cartesian grids for anisotropic and heterogeneous diffusion tensors is extensively treated in~\cite{DaSi2014}.
%%% according to this study, the algorithm~\eqref{eqn:multischeme} is conditionally stable -- as usually, the consistency and stability analysis is given for problems exhibiting sufficient regularity, in order to make possible the application of standard \textit{Taylor expansions}
As a matter of fact, these questions are closely related to the discrete maximum principle for the \textit{Beltrami color flow}, which is typically used in image processing~\cite{DaDi2007}.\\
%%% more references~\cite{DaSo2005}
%%% the following proposition (the proof is omitted) establishes an upper bound for the time-step to ensure numerical stability of the scheme, and a possible proof of this statement (stability and the CFL-condition) is exposed in~\cite{DaSo2005} -- in the general case of $b\neq 0$ and/or $b$ is a space-dependent function, that scheme fails to satisfy the discrete maximum principle and then the \textit{nonnegative method} is considered in~\cite{DaSi2014}
%%% it is worthwhile observing that the transverse components inside anisotropic tensors~\eqref{eqn:anisodiff} may actually occur with any sign and, therefore, their contribution to the diffusion fluxes in~\eqref{eqn:multischeme} is not necessarily directed from higher to lower density concentrations -- in terms of numerical modeling for~\eqref{eqn:multisystem}, this imposes to verify that no unphysical extrema are generated by the approximations, otherwise the discretization could be modified by including high-order limiters~\cite{YaLi2002,ShHa2007,ShHa2011}, for example, to compensate for the presence of spurious concentration gradients
Although the major drawback of explicit algorithms for reaction-diffusion systems like~\eqref{eqn:multischeme} resides precisely in the stringent limitations on the time-step enforced by both the mesh geometry and diffusion rates, through the \textit{CFL-condition}, the parallel implementation using GPUs rehabilitates such methods for the benefits obtained in terms of drastic reduction of the computational time, in contrast to the expensive simulations produced with sequential codes (refer to Section~\ref{sec:performance}).
%%% explicit numerical schemes usually require very small time-steps in order to ensure their stability... the GPU architecture, however, is very well-suited to execute finite difference calculations for many data elements (grid points) simultaneously -- as a consequence, the inconvenience of small time-steps is overcome through parallelization techniques which accomplish thousands of iterations in relatively small wall-clock running times of simulation (a couple of hours)
%%% we adopt the forward Euler scheme for the time discretization because it is directly parallelizable
Moreover, the accuracy in designing spatially compact numerical operators is motivated by the requirements of an optimal parallel implementation~\cite{Fels2008,KlWa2009}, since the nearest-neighbor communication standard is extremely fast with the need for small amounts of local storage in the sub-processors, because only few values are involved to update the numerical solution at each grid node (as shown in Table~\ref{tab:tumorstencil} and Table~\ref{tab:acidstencil}).
%%% and consequently, even very large anisotropic and heterogeneous diffusion tensors become feasible, thanks to an intrinsic parallelism of the methods for conservation laws/models/equations and also the massive number of threads especially in GPU-based computing devices
%%% on that account, the results of this paper encourage undertaking supplementary analysis to include adaptive time-stepping, for example, because Runge-Kutta schemes enjoy explicit error estimations, together with optimized implementation~\cite{Murr2012,RyRo2008} -- and also to extend to systems in the three-dimensional space~\cite{MoCo2010,KuSc2013}
%%% in general, time-implicit or IMEX schemes grow quite computationally inefficient for complex problems and, indeed, high-order Runge-Kutta solvers are important tools for improving the resolution of explicit simulations~\cite{Quar2017,QiHo2013}

\begin{remark}
We point out that the extra-diagonal entries in Table~\ref{tab:acidstencil}, which arise from the anisotropic structure of diffusion tensors~\eqref{eqn:anisodiff}, are indeed responsible for an increased computational complexity, since the matrix-valued version of algorithm~\eqref{eqn:multischeme} does not benefit from sparsity-inducing properties. On the other hand, a different approach to efficient computational arrangements is required for the five-point stencil in Table~\ref{tab:tumorstencil}, because the corresponding matrix-valued scheme is actually nonlinear and need to be reconstructed at each time iteration. Nevertheless, various modern C++ libraries for numerically solving ordinary differential equations are presently available, which are compatible with running on CUDA GPUs through the Thrust interface -- \url{http://thrust.github.io}
%%% more references~\cite{CoPr2011}
\end{remark}

\begin{figure}%[b]
\sidecaption%[t]
\includegraphics[scale=.165]{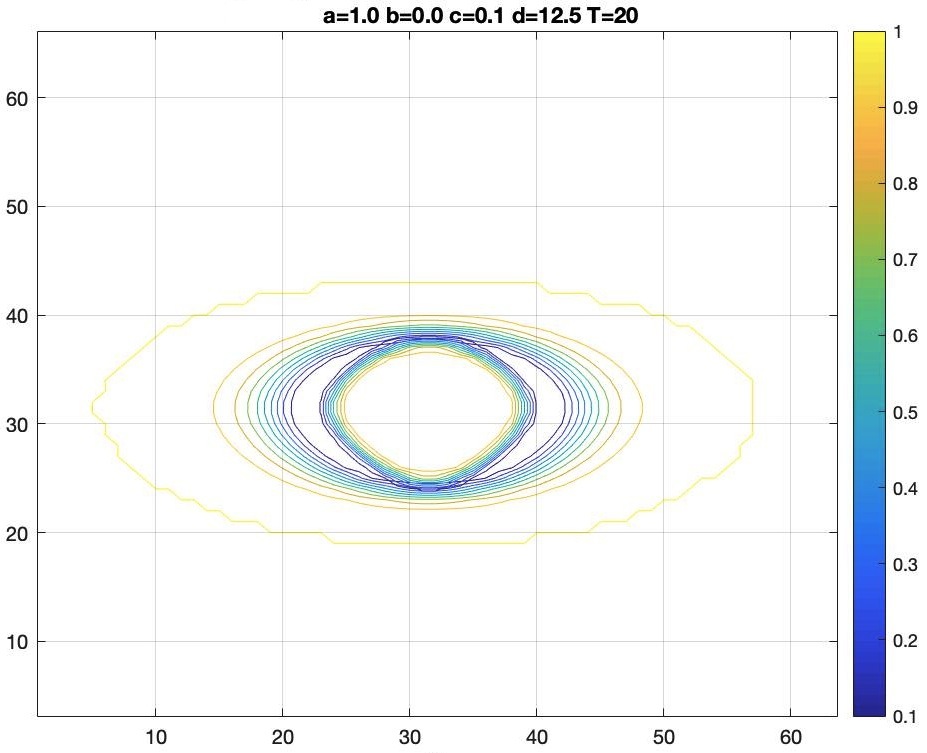}
\qquad \includegraphics[scale=.165]{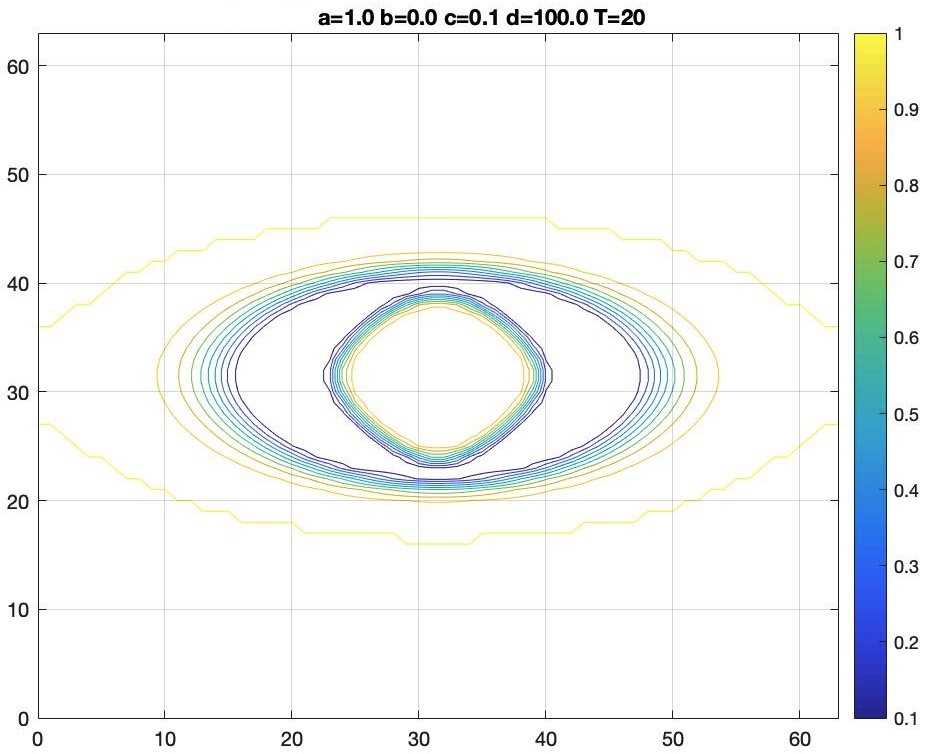}
\caption{Interstitial gap between cancerous and healthy cells for lower (left) and higher (right) values of the aggressiveness parameter $d$ (in the case of lower acid diffusivity along the vertical direction)}
\label{figure2}
\end{figure}

A trademark of the Gatenby-Gawlinski model consists in the capability of reproducing the biologically relevant phenomenon of propagating fronts and, in particular, to predict the development of a \textit{hypocellular interstitial gap}, that is a local region practically depleted of cells where both the healthy and cancerous tissue densities are negligible, which has been experimentally observed~\cite{GaGa1996}.
%%% more references~\cite{FoHo1973,SmGa2007}
%%% traveling wave solutions are discussed in~\cite{GaGa1996,FaHe2009} with some qualitative analysis to reproduce the tumor progression mediated by acidification of the surrounding tissue (the model postulates that an excess of H+ ions is produced by tumor cells as a consequence of their anaerobic, glycolytic metabolism and, in this way, pH is lowered ahead of the advancing tumor front)
%%% for certain parameter values, the healthy tissue could be destroyed prior to the arrival of malignant cells -- more aggressive invasion with a separation zone between the healthy and cancerous cells (i.e. a local region practically depleted of cells located ahead of the advancing tumor front)
In the context of acid-mediated tumor invasion, the dynamical configurations of system~\eqref{eqn:multisystem} with respect to the reaction parameter $d$ essentially exhibit two types of qualitative behavior:
\begin{enumerate}

\item{the \textit{heterogeneous invasion}, occurring for moderate chemical aggression ($d<1$) and characterized by the coexistence of malignant and normal tissues behind the propagating tumor front;}
%%% as shown in Figure~\ref{figurextra3}
%%% the initial data is chosen as ??? and the simulation parameters are ???

%\begin{figure}%[b]
%\sidecaption%[t]
%\includegraphics[scale=.15]{figurextra3}
%\caption{heterogeneous invasion}
%\label{figurextra3}
%\end{figure}

\item{the more aggressive \textit{homogeneous invasion}, which refers to the complete degradation ($d>1$) of local healthy tissue by means of the lactic acid ahead of the propagating tumor front (see Figure~\ref{figure2}(left)); in that case, the emergence of an interstitial gap and its geometrical structure are submitted also to the magnitude of the diffusion rates~\eqref{eqn:anisodiff}, and larger values of $d$ are needed to compensate for lower acid diffusivity in some directions (see Figure~\ref{figure2}(right)).}

\end{enumerate}
%%% from analytical and numerical investigations, the homogeneous invasion turns out to happen faster than the heterogeneous one~\cite{GaGa1996,McGa2014,MoSi2019}, that is in agreement with its more aggressive nature
%%% to estimate the speed of the traveling waves, we have invoked the \textit{LeVeque-Yee formula}~\cite{LeYe1990} whose employing with reaction-diffusion systems and ODEs constitute an original application from~\cite{MoSi2019}

%%% the pictures could be manipulated to make the interstitial gap more evident...

These numerical results suggest that anisotropic and heterogenous acid diffusion is determinant in shaping the propagating tumor fronts, and further investigations are in order for recovering experimental observations about the acid-mediated invasive processes.
%%% namely, the strategy of degradation of the ECM anisotropy and inhomogeneity for enabling the migration of cancer cells through the tissue (as shown in Figure~\ref{figurextra4})
%%% to simulate propagating fronts corresponding to physically relevant situations of interaction involving cancerous and normal species (at the tumor-host interface) because more complicated structures appear also in the case of heterogenous invasion for the choice of an anisotropic, heterogeneous acid diffusion
%%% moreover, we rely on space-averaged wave speed approximation, given by the \textit{LeVeque-Yee formula}~\cite{LeVe1990,MoSi2019}, to quantify the propagation speed of the fronts and, therefore, quantitatively approach the traveling waves solutions...
In this respect, finite volume schemes on staggered grids are particularly well-suited for their ability in accurately reproducing multi-dimensional interfacial problems, especially together with GPU-based computing for biological applications~\cite{MaMi2013,PeMa2019}, and therefore comparable to finite element simulations~\cite{MoSi2019}.
%%% and in fact our simulations correctly reproduce this behavior (i.e. the irregular shape of the interstitial gap) with $x_{i+1/2}$ and $y_{j+1/2}$ located at the initial interfaces between different cell(s) populations for fixed grids

%%% Figure~\ref{figurextra4} shows a stained micrograph of a specimen corresponding to human squamous cell carcinoma (of head and neck) displaying a hypocellular interstitial gap between normal and malignant tissue edges

%\begin{figure}%[b]
%\sidecaption%[t]
%\includegraphics[scale=.55]{figurextra4}
%\caption{experimental data as reported in~\cite{FoHo1973,SmGa2007} -- picture(s) taken from ??? with reprint permission granted}
%\label{figurextra4}
%\end{figure}

%%% to correctly choose values for the anisotropic diffusion matrix one should consider -- 1) the Sylvester criterion for the matrix to be positive definite, and thus define a well-posed parabolic problem (otherwise instabilities appear in the simulations, moreover due to the numerical fail of the positivity condition because of round-off errors) -- 2) the CFL-condition for the global stability of the numerical scheme -- 3) the Neumann-type boundary conditions to avoid spurious effects from the boundary data (even with large computational domains and, therefore, useless large computational times) because also compactly supported initial data immediately become positive over the whole domain for the parabolic operator -- 4) the effects of a Cartesian mesh, which typically reacts incorrectly in the presence of non-cartesian terms of the anisotropic matrices...

%%%%%%%%%%%%%%%%%%%%%%%%%%%%%%

\section{GPU programming and performance evaluation}
\label{sec:performance}

Beside possibly anomalous regularizing effects, a crucial feature for the mathematical modeling of invasion processes with anisotropic and heterogeneous diffusion tensors~\eqref{eqn:anisodiff} is the emergence of strikingly nontrivial patterns~\cite{HiPa2013}.
%%% more references ???
%%% i.e. the formation of irregular structures to which potential invasiveness is associated by exploiting inhomogeneities of the ECM, for example, which can have a profound impact on the rates of cells infiltration, including the formation of characteristic \textit{fingering patterns} (then resulting in morphological instabilities...)
Consequently, the numerical simulations often requires high spatial resolution, and then long-lasting computing execution, for the large amount of data to be traded in order to accurately capture the details of biological phenomena. Moreover, especially for clinical practitioners and applied scientists involved in setting up realistic experiments, the possibility of running fast comparative simulations using simple algorithms implemented into affordable processors is of primary interest.

%\begin{figure}%[b]
%\sidecaption%[t]
%\includegraphics[scale=.35]{figurextra5}
%\caption{irregular shapes produced by exploiting ECM inhomogeneities~\cite{AnWe2006,Ande2007} -- picture(s) taken from ??? with reprint permission granted}
%\label{figurextra5}
%\end{figure}

Parallel computing based on modern Graphics Processing Units (GPUs) has the advantage of high performance at a relatively low energetically and monetary costs.\\
%%% nevertheless, the overall benefits of parallel processing are slightly more complicated than just decreasing the running time of some programs~\cite{MaNi2018}
%%% already in 2002, commodity graphics cards started to outperform Central Processing Units (CPUs) -- as GPUs grew faster and cheaper, the interest to harvest their power for applications others than graphical display originated, around 2006, what is known as General Purpose GPU computation (GPGPU) -- https://hgpu.org -- by the year 2009, GPUs that could be bought out the shelf had a theoretical peak performance of more than a thousand single precision GFLOPs ($10^9$ floating point operations per second) that is almost ten times more than their multi-core CPU counterpart -- nowadays, GPUs found in personal computers and laptops can perform double precision computations with a ratio of speed over cost larger than any other parallel computing architecture and, additionally, GPUs are also energetically efficient making them an affordable and portable option for parallel computation
The codes used in this article are programmed using the NVIDIA$^{\textcopyright}$ Compute Unified Device Architecture (CUDA) platform, which is designed to support GPUs execution for data parallelization~\cite{CUDA2019}. Through the CUDA implementation, graphics cards are programmed with a medium-level language, which is recognized as an extension to C/C++, without requiring sophisticated hardware expertise~\cite{KiHw2016,SaKa2010}.\\
%%% we refer to~\cite{KiHw2016,SaKa2010} for a comprehensive introduction to General Purpose GPU (GPGPU) parallel computing, which include details about the CUDA-based programming and the architecture of current generation graphics cards.
All simulations\footnote{The code for reproducing the numerical tests is available upon request to the authors.} in this article are performed using NVIDIA$^{\textcopyright}$ graphics cards GTX 1080 with 2560 CUDA cores and 8 Gb RAM, installed on a processor HP DL585G7 4 AMD Opteron 6128 with 8 cores, clock frequency 2.0 GHz, 64 Gb RAM, operating system Linux centOS 6.5 amd64, compiler GNU gcc 4.4.7 and NVIDIA$^{\tiny{\textcopyright}}$ CUDA 9.1 Linux 64 bit toolbox. For a comparative study of computing performance, the same algorithms are implemented serially on a single CPU processor HP DL585G7 4 AMD Opteron 6128 CPU with 8 cores.
%%% the same numerical scheme and parameter values are used for both implementations

%\begin{figure}%[b]
%\sidecaption%[t]
%\includegraphics[scale=.55]{figurextra6}
%\caption{Compute capability of the NVIDIA's GeForce GTX 1080 -- https://www.nvidia.com/en-us/geforce/products/10series/geforce-gtx-1080/}
%\label{figurextra6}
%\end{figure}

The standard logical steps for implementing the parallel codes are as follows:
\begin{enumerate}

\item{the initial data and numerical parameters are loaded into the CPU (host) memory, and then transferred to the GPU (device) storage, namely the \textit{global memory}};
\item{at each time iteration, the GPU provides massive computing activities, by executing in parallel across the sub-processors the algorithm for the spatial discretization, since the current unknowns only depend on those already computed at the previous iteration};
\item{the GPU uploads the local simulation results to the CPU memory};
\item{the CPU updates the value of the time-step, according to the constraints imposed to guarantee the numerical stability, and then it restarts/stops the parallel computing process.}

\end{enumerate}

\begin{figure}%[b]
\sidecaption%[t]
\includegraphics[scale=.30]{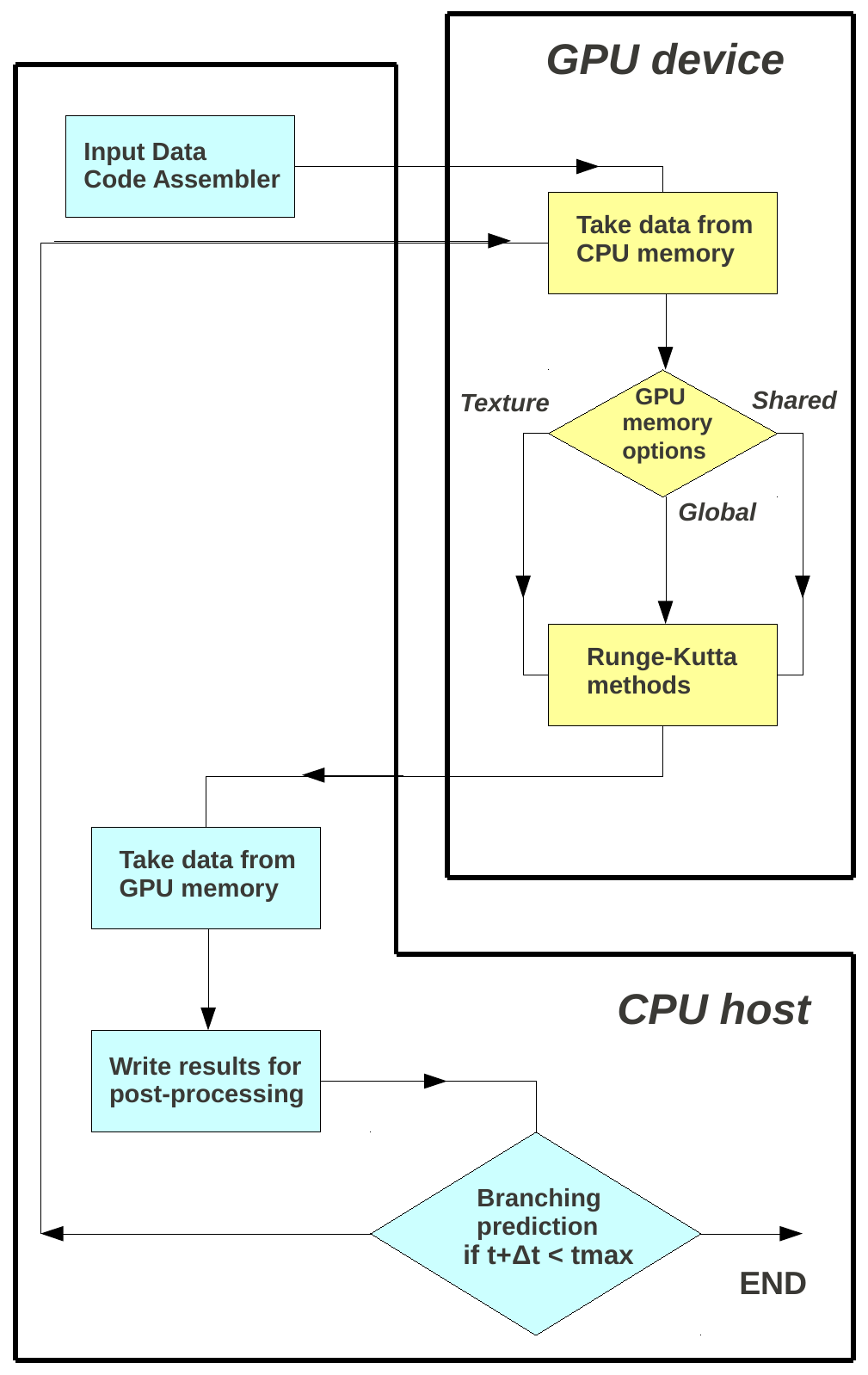} 
\quad \includegraphics[scale=.45]{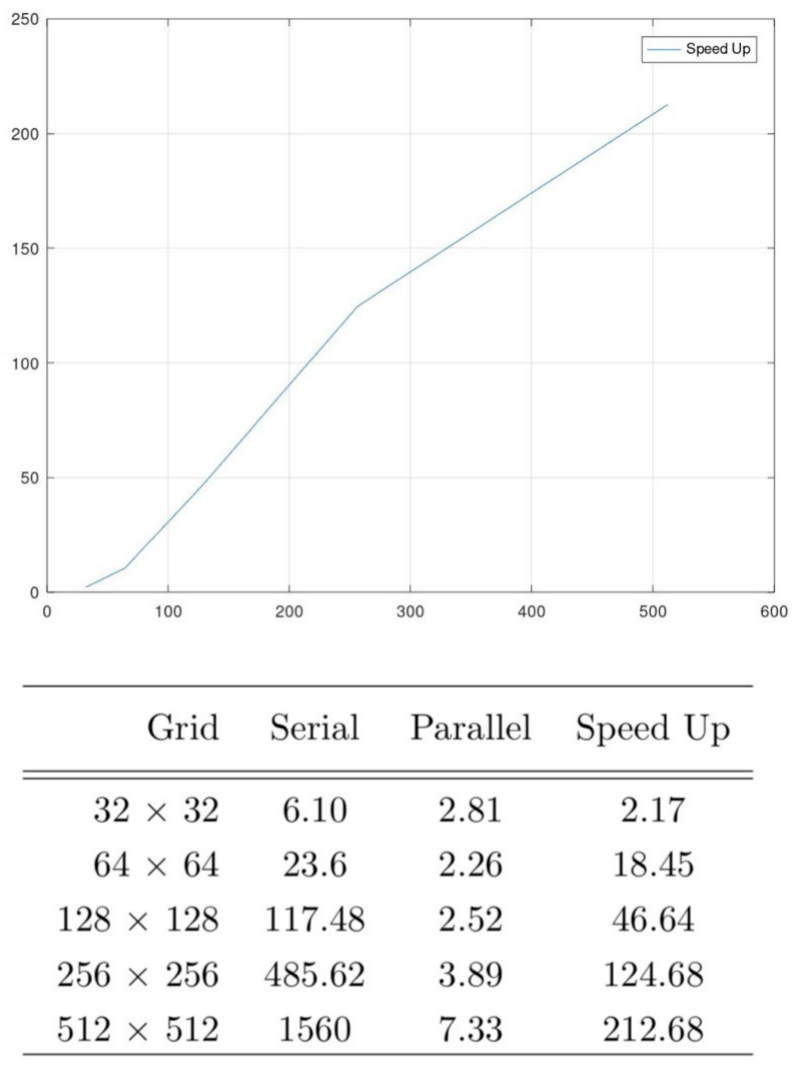}
\caption{General steps of GPU-based parallel codes (left) and \textit{speed-up} evaluation (right)}
\label{figure3}
\end{figure}

\noindent From the point of view of programming, this approach leads to designate an external \textit{loop} for the time integration, whereas the spatial approximation at each iteration is carried out in parallel by the GPU within a \textit{master-slave} model, which is sketched in Figure~\ref{figure3}(left). Finally, the performance comparison between serial and parallel execution for the above simulations is shown in Figure~\ref{figure3}(right), where the \textit{speed-up} is illustrated as a function of the mesh size, for instance.
%%% where the CPU (host) is the owner of time clock activities, and the GPU (device) is the owner of the (parallel) massive computing activities related to the spatial discretization, moreover the CPU is also the master because it controls the parallel executions on GPUs

The CUDA implementation of the numerical codes is currently undertaking further optimizations, for example by employing other GPU \textit{local memories} for achieving \textit{dynamic parallelism}, and these issues are oriented to the three-dimensional modeling of experimental data.

%\begin{programcode}{Program Code}
%\begin{verbatim} to reproduce parts of numerical codes \end{verbatim}
%\end{programcode}

%\begin{svgraybox}
%to emphasize complete paragraphs
%\end{svgraybox}

%%%%%%%%%%%%%%%%%%%%%%%%%%%%%%

\begin{acknowledgement}

This work is supported by the French government, managed by the ANR under the UCA JEDI Investments for the Future project, reference n. ANR-15-IDEX-01. All numerical simulations are performed on the Linux HPC parallel cluster Caliban -- \url{https://caliband.disim.univaq.it} -- at the University of L'Aquila (Italy). This article is greatly motivated by fruitful discussions with the Systems Biology Group -- \url{https://sbglab.org} -- at the Department of Experimental Medicine, Sapienza University of Rome (Italy).
%%% within the ??? project Phase Transitions in Biology through Mathematical Modelling

\end{acknowledgement}

%\biblstarthook{}

\end{document}